\documentclass[letter]{aa} 
\usepackage{graphicx,natbib}
\usepackage{inputenc}
\usepackage{amssymb}

\usepackage{txfonts}
\usepackage{color}

\begin{document} 

   \authorrunning{Charbonnel et al.}
   \titlerunning{Are there any first-generation stars in globular clusters today?}

   \title{Are there any first-generation stars in globular clusters today?}




   \subtitle{}

   \author{Corinne Charbonnel\inst{1,2}, William Chantereau\inst{1}, Martin Krause\inst{3,4}, Francesca Primas\inst{5}, Yue Wang\inst{5,6}
          }

   \institute{Department of Astronomy, University of Geneva, Chemin des Maillettes 51, 1290 Versoix, Switzerland\\
   \email{corinne.charbonnel@unige.ch} \and
   IRAP, CNRS UMR 5277, Universit\'e de Toulouse, 14, Av. E.Belin, 31400 Toulouse, France \and 
   Max-Planck-Institut f\"ur Extraterrestrische Physik, Giessenbachstr. 1, 85741, Garching, Germany \and
   Excellence Cluster Universe, Technische Universit\"at M\"unchen, Boltzmannstrasse 2, 85748, Garching, Germany \and
   European Southern Observatory, Garching, Germany \and
   Key Laboratory of Optical Astronomy, National Astronomical Observatories, Chinese Academy of Sciences, China \\
             }

   \date{Received August 14, 2014; accepted September 1, 2014; A\&A 569, L6}

 
  \abstract
   {Several models compete to explain the abundance properties of stellar populations in globular clusters. 
   One of the main constraints is the present-day ratio of first- and second-generation stars that are currently identified based on their sodium content.}
   {We propose an alternative interpretation of the observed sodium distribution, and suggest that stars with low sodium abundance that are counted as members of the first stellar generation could actually be second-generation stars.}
  {We compute the number ratio of second-generation stars along the Na distribution following the fast rotating massive star model using the same constraints from the well-documented case of NGC 6752 as in our previous developments.}
  {We reproduce the typical percentage of low-sodium stars usually classified as first-generation  stars
  by invoking only secondary star formation from material ejected by massive stars and mixed with original globular cluster material in proportions that account for the Li-Na anti-correlation in this cluster.}
  {Globular clusters could be totally devoid of first-generation low-mass stars today. This can be tested with the determination of the carbon isotopic ratio and nitrogen abundance in turn-off globular cluster stars. Consequences and related issues are briefly discussed.
  }
    
 \keywords{globular clusters: general}
\maketitle

%

\section{Introduction}
\label{Section:Intro}
The well-documented O-Na anti-correlation is now accepted 
as the main chemical characteristic of stellar populations in bona fide globular clusters (GC) both in our Galaxy and in the Local Group (e.g. \citealt{carretta09a,Mucciarellietal09,Larsenetal14} and references therein).
In these massive and old star clusters, long-lived low-mass stars (LMS) exhibit large and anti-correlated spreads in Na and O abundances.
This pattern has not yet been found among field stars covering the same metallicity range as Galactic GCs (but see \citealt{Carretta10} who suggest that $\sim 1.4 \%$ of the field metal-poor stars are likely Na-rich stars evaporated from GCs).
It is interpreted as the presence of (at least) two stellar generations in every individual GC. 
First-generation (1G) stars are defined as those that have Na and O abundances similar to that of halo field stars of similar metallicity. On the other hand, second generation (2G) stars are identified thanks to their Na overabundances and (eventual) O depletion; they are expected to have formed out of the ashes of hot hydrogen burning ejected by more massive, fast-evolving 1G GC stars and mixed with original proto-cluster gas \citep[e.g][]{Prantzos07}. 

The present-day 1G/2G ratio is only estimated based on abundance criteria \citep{Prantzos06}. 
Considering abundance determination uncertainties, \citet{carretta09b} labelled the 1G stars 
of individual GCs as those characterized by [Na/Fe] ratios falling in the range between [Na/Fe]$_{min}$ and [Na/Fe]$_{min}$+0.3~dex\footnote{This corresponds to $\sim 4 \sigma$([Na/Fe]), where $\sigma$([Na/Fe]) is the star-to-star error on [Na/Fe] in each individual GC.}, 
where [Na/Fe]$_{min}$ is the lowest Na abundance derived in each specific GC. 
All the other GC stars departing from this low Na 
area are considered 2G stars\footnote{\citet{carretta09b} further divided the 2G group into an ``intermediate" and an ``extreme" components (the latter not being present in all the clusters), depending on their O depletion with respect to the highest O abundance observed, i.e., on the [O/Na] ratio.}.
Using their homogeneous spectroscopic study of $\sim$ 1400 red giants in 15 Galactic GCs, they 
showed that the 
1G component is present at a constant level of $33 \pm 1 \%$ in all the Galactic GCs surveyed so far (see also \citealt{Carretta13}).

This ratio is a key 
parameter for the scenarii of secondary star formation that aim at explaining the presence of multiple populations and the observed abundance patterns in GCs. 
It constrains the initial GC mass depending on the invoked 1G polluters, namely the fast rotating massive stars (25-120 $M_{\odot}$, hereafter FRMS; \citealt{Maeder06}; \citealt{Prantzos06}; \citealt{Decressin07a}; 
\citealt{Krause13}), the massive asymptotic giant branch stars (6-11 $M_{\odot}$, AGB; \citealt{Ventura01,Ventura13,D'ercole10,Ventura11}), as well as the possible contribution of massive binary stars \citep{DeMink09,Izzard13}, 
and of FRMS paired with AGB stars \citep{Sills10} or with high-mass interactive binaries \citep{Bastianetal2013a,CassisiSalaris2014}. 
Whatever the actual polluting stars are and assuming ``classical" values for the initial mass function (IMF), the observed value of this ratio ($33 \pm 1 \%$) implies that more than $95 \%$ of the 1G LMS were ejected from the GCs 
whose initial stellar masses were 8 - 25 times larger than 
today \citep{Prantzos06,Decressin07b,Decressin10b,Carretta10,D'ercole10,SchaererCharbonnel2011}.
Fast gas expulsion has been suggested to solve this so-called mass-budget problem, such that most of the 1G stars would be lost with the gas (e.g. \citealt{Decressin10b}). 
However, stellar and supernovae feedback is unlikely to accomplish this (accretion onto dark remnants might work however, \citealt{Krause12}), and 
constraints from dwarf galaxies limit the amount of halo stars with GC metallicities (e.g. \citealt{Larsenetal14}).

Here we show that in the FRMS scenario, a large fraction of 2G stars can form with Na and O abundances similar to those of the stars that are presently counted as 1G stars (i.e., low Na and high O similar to that of halo field stars; \S\ref{Section:abundpatternspredicted}). 
We propose observations of the carbon isotopic ratio in turn-off GC stars as a critical test to discriminate between ``true" 1G LMS formed from pure original proto-cluster gas and ``fake" 1G LMS actually made of a mixture of massive star ejecta and original material, and briefly discuss the case of the other polluter scenarii (\S\ref{Section:predictionsforfutureobservations}).
If confirmed, this result would lead to a new paradigm shift in the domain, by considerably alleviating the mass budget problem, and reconciling GC issues with constraints provided by young massive clusters (YMC) and dwarf galaxies (\S\ref{Section:Discussion}). 

\section{Theoretical sodium distribution}
\label{Section:abundpatternspredicted}

\subsection{FRMS guidelines}
\label{subsection:FRMSguidelines}

The 1G is supposed to form from proto-cluster gas that was already enriched in heavy metals (i.e., iron-group, alpha-, and neutron-capture elements) during the Galactic chemical evolution, 
and it has an initial (or original) composition similar to that of field stars of similar [Fe/H]. 
In particular, the lowest Na abundance  [Na/Fe]$_{min}$ (see \S~\ref{Section:Intro}) as well as the highest O abundance observed in a given GC 
are considered as the original abundances of these elements in the proto-GC gas; the same applies to all the light elements whose abundances are correlated or anti-correlated with that of Na, namely Li, F, C, N, Al, Mg.

Very early in GC evolution after the formation of the 1G, photo-dissociation of molecular hydrogen by Lyman-Werner photons emitted by 1G massive stars is expected to prevent further ``classical" star formation from occurring \citep{ConroySpergel2011,Krause13}.
However, in the FRMS scenario \citep{Prantzos06,Decressin07a,Decressin07b,Krause13}, the formation of 2G LMS is expected to occur in the immediate vicinity of polluter stars with initial masses higher than $\simeq$ 25~M$_{\odot}$.
More precisely, if 1G massive stars have mechanical ejections at or near rotation breakup, massive and gravitationally unstable equatorial discs can form and be fed both by H-burning products ejected by FRMS and by original proto-GC gas. 
The episode of 2G star formation is expected to last over a period of $\simeq$ 3.5 - 8.8~Myr after the formation of the 1G massive stars, the exact duration depending on the upper mass limit for stars to explode as supernovae or to become black holes. 

If substantial amounts of gas were not converted into stars or accreted towards the FRMS when the first energetic SNe explode, this gas should remain in the GC (feedback energy would only suffice to stir turbulence below escape speed) and would likely mix with the SNe ejecta. Because there is no hint for this to occur from LMS in most GCs, more energetic feedback, for example due to accretion onto dark remnants, needs to clear out this gas later, but before it can form stars again. Recent observations suggest even that YMC keep their gas only a few Myr after star formation (\citealt{Smithetal06,Bastianetal2013a,Bastianetal2014,CabreraZirietal2014,BastianStrader2014}; Hollyhead et al. 2014, in prep.). 
This could mean that either the feedback energy has been underestimated, or that at least all the strongly gravitationally bound gas is converted into stars. In any case, the timescale of a few Myr after the onset of star formation for gas to be present would still be sufficient for the FRMS ejecta to encounter and mix with such gas, as required by the lithium constraint (see below).

\subsection{Sodium abundance distribution of the second stellar generation. Qualitative description}
\label{subsection:qualitative}

The material ejected 
by the FRMS presents various degrees of enrichment in H-burning products as the polluters evolve. 
The polluter stars first release material of original composition. 
Actually, at the very beginning of the main sequence evolution, the stellar ejecta of the massive stars still contain fragile elements like lithium, beryllium, boron, and fluorine, which are protected from proton-captures in the most external and coolest stellar layers\footnote{Typically, stellar models with Z=0.0004 and initial masses of 40, 60, and 120~M$_{\odot}$ do preserve Li in their most external 0.02, 0.018, and 0.036~M$_{\odot}$ (Ekstr\"{o}m, Georgy, \& Groh, private communication).}.
Then LiBeBF-free material is released but still with Na and O abundances similar to the original abundances. 
Finally, as stars evolve along the main sequence and the luminous blue variable (LBV) phase, the products of the CNO-cycle and of the NeNa- and MgAl-chains are transported 
from the core towards the stellar surface and within the slow wind thanks to rotation-induced mixing. 
Importanty, the presence of the fragile Li and Be \citep{Pasquini05,Pasquini07} in the long-lived LMS enriched in Na indicates that the polluters ejecta did mix to various degrees with pristine material to form 
the 2G stars.
In summary, one expects to first form 2G stars with Na and O contents similar to that of the 1G (i.e., ``fake" 1G stars), and then 2G stars with various degrees of Na enrichment and O depletion. 
Note that the maximum time spread of this sequence corresponds typically to the main sequence lifetime of the massive polluters (i.e., 3.5 and 8.8~Myr for the 120 and 25~M$_{\odot}$ stars with [Fe/H]=-1.56 respectively).

\subsection{Quantitative sodium abundance distribution}
\label{subsection:quantitative}
To compute the number fraction of 2G LMS born with a specific chemical composition, we follow the method and the assumptions presented in \citet{Decressin07b}.
We use the time-dependent ejecta of \citet{Decressin07a} models for FRMS computed at the metallicity of NGC 6752 ([Fe/H]=-1.56) and assume that 100$\%$ of the H-processed ejecta are released in slow winds and recycled into 2G. The behaviour of Li with respect to Na observed in this well-studied GC is used 
to estimate the dilution factor between the material ejected in the slow equatorial stellar winds and original interstellar matter.  
The amount of material that is made available to form 2G stars is very large.
Indeed, \citet{Decressin07a} FRMS models lose about one third to half of their initial mass 
along the main sequence and the LBV phase, 
and roughly the same amount of GC gas of original composition is required to account for the observed Li-Na anti-correlation in NGC~6752. 

\begin{figure}
 \center{
\includegraphics[width=\columnwidth,height=0.6\columnwidth]{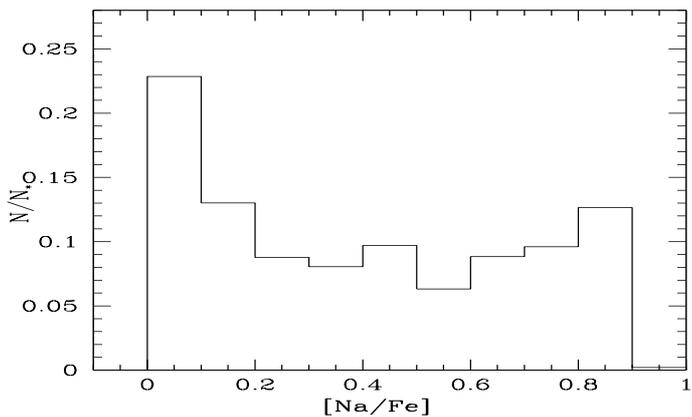}
}
\caption{
Theoretical distribution of the sodium abundance in second-generation low-mass stars at birth (see text for details) 
}
\label{Figure:SodiumNitrogenDistribution}
\end{figure}

Figure~\ref{Figure:SodiumNitrogenDistribution} shows the prediction for the number fraction of 2G stars born with different Na abundances when considering \citet{Salpeter55} IMF for the FRMS polluters of masses higher or equal to 25~M$_{\odot}$, and assuming that the 2G consists only of stars with initial masses below or equal to 0.8~M$_\odot$ formed with \citet{Paresce00} log-normal mass distribution all along the Na distribution.
The dilution factor $\it{a}$ (Eq.27 of \citealt{Decressin07b}) varies here between 0.95 and ${\it{a_{min}}}=0.2$, and we take [Na/Fe]$_{min}=0$. 
The corresponding initial helium abundance distribution successfully explains the lack of Na-rich giants in NGC 6752 \citep{Charbonnel13}.
As expected, numerous 2G stars can form with Na abundances that perfectly overlap that of the stars that are currently considered 1G stars (i.e., with low Na).
Quantitatively, the theoretical proportion of 2G stars with [Na/Fe] between 0 and 0.3~dex, which is the domain generally considered that of 1G stars (\S\ref{Section:Intro}), is $45 \%$. 
For a dilution factor varying between a=0.8 (instead of 0.95) and keeping ${\it{a_{min}}}=0.2$, we obtain $34 \%$ of 2G stars with [Na/Fe] between 0 and 0.3~dex.

This result is in striking agreement with the percentage of 1G stars obtained when using the observed [Na/Fe] 
distribution, i.e., $33 \pm 1 \%$ (\S~\ref{Section:Intro}). 
Therefore, one can in principle account for the whole Na abundance range observed today for long-lived LMS by invoking only secondary stellar formation 
in the neighborhood of FRMS. 
In other words, GCs could be totally devoid of 1G LMS, and all the LMS we observe today could have formed out of the ejecta of 1G massive stars mixed with original gas in roughly 50 -- 50 $\%$ proportions.
This requires that 100$\%$ of the H-processed material released by the FRMS stays within the GC and is recycled into the 2G; on the contrary, if 
the FRMS material with low Na content is ejected from the GC or not recycled, the stars located in the low Na region of the O-Na anti-correlation would be ``true" 1G stars.




\section{Observational tests and differences with other scenarii}
\label{Section:predictionsforfutureobservations}
We propose two observational tests to disentangle between ``true" 1G LMS formed from pure original proto-cluster gas and ``fake" 1G LMS made of a mixture of 1G massive star ejecta and original material.

\subsection{Carbon isotopic ratio}
\label{subsection:cir}
Most of the matter ejected in the slow winds of the FRMS is near CN-equilibrium, i.e. its carbon isotopic ratio (hereafter $^{12}$C/$^{13}$C) is $\sim$ 3.8. 
When considering dilution with original gas that has much higher $^{12}$C/$^{13}$C (here we take the solar value of 90, but the results are consistent with those of \citealt{Decressin07b}, who assumes an initial value of 240; see also  \citealt{Chiappinietal08}), one 
obtains intermediate values. 
Figure~\ref{Figure:CarbonIsotopicRatiovsSodium} shows the predicted variation of $^{12}$C/$^{13}$C as a function of the Na abundance in the initial composition of 2G stars, under the same assumptions made in \S~\ref{subsection:quantitative}.
Only the 2G stars with the lowest Na abundances are predicted to have high $^{12}$C/$^{13}$C, close to the original ratio at birth.
 Initial $^{12}$C/$^{13}$C=20 is already predicted for [Na/Fe]$_{min}+0.05$dex, and all the 2G stars with [Na/Fe] higher than $\sim$ [Na/Fe]$_{min}+0.15$~dex are expected to be born with $^{12}$C/$^{13}$C lower than 15. 
In contrast, 1G stars with [Na/Fe] between 0 and 0.3 dex would be born with the original high $^{12}$C/$^{13}$C. 

The determination of $^{12}$C/$^{13}$C in GC stars in the domain between [Na/Fe]$_{min}$ and [Na/Fe]$_{min}+0.3$dex thus appears to be a powerful observational way to distinguish between ``true" and ``fake" 1G LMS. 
Note, however, that once LMS evolve towards the red giant branch, their surface $^{12}$C/$^{13}$C decreases from its original value due to dilution of H-processed core material with their external layers (the so-called first dredge-up). 
Later on, this quantity also decreases when the stars reach the RGB bump, probably due to thermohaline mixing \citep[see e.g.][]{CharbonnelBW98,Gratton00,CCJPZ07,CharbonnelLagarde10}\footnote{A 0.8M$_{\odot}$ model with [Fe/H]=-1.75 has a $^{12}$C/$^{13}$C of 40 and of 7 (starting from initial 90) respectively after the first dredge-up completion and at the RGB tip, in agreement with observations for field RGB star (Chantereau et al. in prep.; see also \citealt{CCJPZ07}).}.
Therefore, the suggested observational test has to be performed in stars before the occurrence of the first dredge-up, i.e. in turn-off stars or in only slightly evolved sub-giant stars.

$^{12}$C/$^{13}$C has actually been determined for a handful of subgiant stars that span a relatively broad range of Na abundance in NGC~6752 and 47Tuc \citep{Carretta05}. 
The observed  $^{12}$C/$^{13}$C values range between 3 and 12. 
According to their position in the GC colour-magnitude diagrams, these stars should not have fully undergone first dredge-up, and must be born with these low $^{12}$C/$^{13}$C. 
Among them, only a couple have relatively low Na, but the stars with the lowest Na actually have a $^{12}$C/$^{13}$C of the order of 9 -- 12. 
This is in very good agreement with the predictions shown in Fig.\ref{Figure:CarbonIsotopicRatiovsSodium}. 
However, the data are sparse and observations in turn-off stars are urgently needed.  


\subsection{Nitrogen abundance}
\label{subsection:Nitrogen}

\begin{figure}
\center{
\includegraphics[width=0.8\columnwidth]{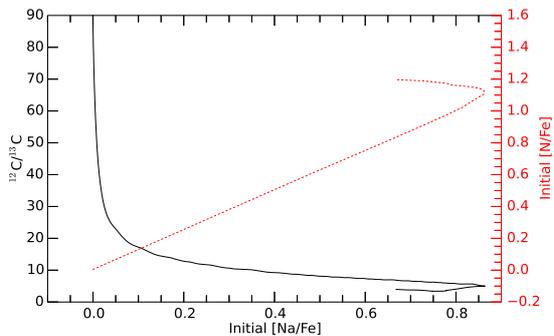}
}
\caption{
Predicted carbon isotopic ratio and nitrogen abundance (solid black and dotted red lines, respectively) as a function of sodium abundance in the second-generation stars at birth
}
\label{Figure:CarbonIsotopicRatiovsSodium}
\end{figure}

As shown in Fig.~\ref{Figure:CarbonIsotopicRatiovsSodium}, the initial nitrogen abundance of 2G stars is also expected to vary strongly with Na, again due to contamination of the original gas (here we assume initial [N/Fe] = 0) with CN-processed material at the equilibrium.
Indeed, [N/Fe] varies by 0.4 dex when [Na/Fe] varies between [Na/Fe]$_{min}$ and [Na/Fe]$_{min} +0.3$ dex.
Therefore, the determination of nitrogen abundances could also provide, in principle, a good observational test. 
Again, turn-off stars are the ideal targets, to avoid surface abundance changes due to first dredge-up and thermohaline mixing as the stars evolve\footnote{In a 0.8M$_{\odot}$ stellar model with [Fe/H]=-1.75, surface [N/Fe] increases by $\sim$ 0.02~dex due to first dredge-up, and $\sim$ 0.6 dex due to thermohaline mixing (Chantereau et al. in prep.)}. 

In the case of NGC 6752, the observational situation is unclear. 
For stars spanning the range between [Na/Fe]$_{min}$ and [Na/Fe]$_{min} +0.3$, \citet{Carretta05} and \citet{Yongetal08}, find N abundance variations of 0.5 and 1 dex respectively. 
The difference between the two analyses cannot be attributed to internal variations due to first dredge-up nor thermohaline mixing in the two samples, as the lowest [N/Fe] value determined by   \citet{Carretta05} is considerably higher than that of \citet{Yongetal08}. 
Rather, \citet{Yongetal08} speculate that the differences in the observed [N/Fe] distribution come mainly from the difficulties in determining accurate N abundances from CN lines. We urge observers to perform new analyses of N in turn-off GC stars.

\subsection{Expectations from the other polluter scenarii}
\label{subsection:otherscenarii}
To alleviate the mass budget problem by enlarging the mass domain of the polluters, \citet{DeMink09} proposed massive binary stars ($\sim$ 9 to 20~M$_{\odot}$) as an additional source of H-burning ashes. 
However, about 1/3 of the material released by these stars is relatively unprocessed with a mass fraction of Na in the raw ejecta lower than twice the original value (their Fig.1), and this material is also expected to mix fifty-fifty with original gas. 
 Therefore, the formation of  ``fake" 1G stars is also expected in this case, but their number would be $\sim 4$ times higher than in the FRMS case when considering the mass-weighted IMF (their Fig.2). This is clearly excluded by the observations.

In the AGB scenario, the O-Na anti-correlation can be obtained only through a complicated dilution model, as the AGB yields do actually produce a correlation between O and Na \citep[e.g.][ and references therein]{Ventura13}.
Secondary star formation is expected to start after the ejection of the gas and of the SNe ejecta, about $\sim 50 - 100$~Myr after the formation of the 1G.
The sequence of events is that ``pure" 2G stars form first from raw AGB ejecta, until pristine gas is re-accreted and falls back into the GC core regions, mixes with the AGB winds, and forms 2G stars with diluted ejecta \citep{D'ercole08,D'ercole10,D'ercole12}. In this case, the stars observed in the domain between [Na/Fe]$_{min}$ and [Na/Fe]$_{min}+0.3$dex  are expected to be ``true" 1G stars with original $^{12}$C/$^{13}$C and N abundance.
Therefore, the observational tests we propose will also be able to discriminate between the AGB and the FRMS scenarii. 
 
 

\section{Summary, consequences, and open issues}
\label{Section:Discussion}

We show that the whole Na range exhibited today by long-lived, low-mass GC stars, as well as the percentages of the so-called 1st and 2d stellar generations, can be explained by invoking secondary star formation from FRMS ejecta mixed with pristine material. Therefore, GCs could be totally devoid of 1G LMS today. We propose observational tests to discriminate between true and fake 1G stars and to constrain the various polluter scenarii.

If the absence of ``true" 1G stars in GCs today is confirmed, we should find out whether all 1G LMS were lost from the GCs or whether they have not formed, and why. 
The first case seems very improbable in view of the mass budget and gas expulsion problems, which would be even more exacerbated. 

In the second case, the mass initially locked in 1G GC massive stars could have been only two to four times the present-day stellar mass, since roughly one third to half of the FRMS mass 
is made available for recycling into the 2G after having been mixed with original gas in $\sim$ fifty-fifty proportions. Of course, this is a minimum value for the initial GC mass as we assume that $100 \%$ of the available material is recycled into LMS. 
But this would definitively release the current tension between the different model predictions and the constraints coming from dwarf galaxies. 
We should then understand why LMS could not form initially. Clouds at higher temperature, which may occur after formation of the first massive stars, are known to prevent LMS formation (\citealt{Klessenetal07}, sharp turn-down below 7 M$_{\odot}$). Magnetic fields, radiation feedback, and a steeper initial density profile have a similar effect \citep{Girichidisetal11,Petersetal11}. 
There are also observational hints that star-forming regions with higher stellar density have a more top-heavy mass function (e.g. \citealt{Kryukovaetal14}). 
Work is in progress on all these open issues. 



\begin{acknowledgements}
This research was supported the Swiss National Science Foundation (FNS), the International Space Science Institute (International Team 271 ``Massive Star Cluster across the Hubble Time"), the cluster of excellence ``Origin and Structure of the Universe" (http://www.universe-cluster.de), and the Centre National de la Recherche Scientifique (CNRS).
\end{acknowledgements}

\bibliographystyle{aa}
\bibliography{Bibliography}

\end{document}